\documentclass[preprint,aps]{revtex4}
\newcommand{\be}{\begin{equation}}
\newcommand{\ee}{\end{equation}}
\newcommand{\ba}{\begin{eqnarray}}
\newcommand{\ea}{\end{eqnarray}}

\begin{document}

\draft

\title{Relativistic Electrodynamics \\ 
       of Spinning Compact Objects}

\author{Hongsu Kim\footnote{e-mail: hongsu@kasi.re.kr}}

\address{International Center for Astrophysics, Korea Astronomy and Space Science Institute, Daejeon 305-348, KOREA}

\author{Hyung Mok Lee\footnote{e-mail: hmlee@astro.snu.ac.kr}}

\address{Astronomy Program, School of Earth and Environmental Sciences, \\ 
Seoul National University, Seoul 151-742, KOREA}

\author{Chul H. Lee\footnote{e-mail: chulhoon@hanyang.ac.kr} and
Hyun Kyu Lee\footnote{e-mail: hyunkyu@hanyang.ac.kr}}

\address{Department of Physics, Hanyang University, Seoul 133-791, KOREA}


\begin{abstract}
A theoretical study of some electrodynamic features of a region close to a {\it slowly-rotating} 
magnetized relativistic star is performed. To be a little more specific,
based on the solution-generating method given by Wald, the magnetic fields around
both uncharged and (slightly) charged relativistic stars have been obtained. 
Particularly for a charged relativistic star, again following the 
argument by Wald, the star was shown to gradually accrete charge until it reached an equilibrium 
value $\tilde{Q}=2B_{0}J$. This value of the equilibrium charge seems to be generic as a rotating
black hole is known to accrete exactly the same amount. Although these results are 
equally relevant to all species of slowly-rotating relativistic stars, we particularly have the 
rotating neutron star in mind. As such, it would be of some interest to attempt to make contact with 
a real pulsar case. Thus, we discuss how many of the theoretical results obtained in the
present work can be carried over to a realistic, general relativistic description of a pulsar's
magnetosphere. 
\end{abstract}


\maketitle

\narrowtext

\newpage
\begin{center}
{\rm\bf I. INTRODUCTION}
\end{center}

In the present work, we perform a theoretical study of some electrodynamic features of
a region close to a {\it slowly-rotating} relativistic star. Here, by relativistic star, we mean
a highly self-gravitating compact object. Particularly, we have a rotating neutron star
in mind. Thus, we first begin by providing the rationale for treating the vicinity of a spinning compact
object, such as a rotating neutron star, relativistically as a non-trivial curved spacetime. 
Since the pioneering proposals
by Gold \cite{gold} and by Pacini \cite{pacini}, it has by now been widely accepted that, indeed, pulsars
might be rotating magnetized neutron stars. Nevertheless, since then, nearly all studies 
on pulsar electrodynamics have been performed by simply treating the region surrounding
the rotating neutron stars as flat. This simplification may be sufficient just to gain
a rough understanding of the origin of the pulsars' radiation. However, nearly  thirty years
of studying pulsar electrodynamics have passed, and 
now, it seems we should look into the situation in a more careful and rigorous manner. 

In the present work,
we take one step forward in this direction. We particularly aim at determining
the amount of charge on and the structure of the magnetic field 
around a slowly-rotating neutron star by treating its exterior region, literally, as a curved spacetime 
and employing the algorithm of Wald \cite{wald}. Indeed, thus far, only a
handful of earlier studies concerning the electrodynamics of rotating neutron stars have appeared in 
the literature, and they include the works by Muslimov et al.
\cite{muslimov} and by Prasanna and Gupta \cite{muslimov}. Although their motivation, approach, and
tools are completely different from those of our present work, Rezzolla et al. \cite{rezzolla}
also considered analytic solutions of Maxwell equations in the internal and external regions of a
slowly-rotating magnetized neutron star. \\
Indeed, we now
have a considerable amount of observed data for various species from radio pulsars \cite{gj} to
(anomalous) X-ray pulsars \cite{x-ray}. Even if we take the oldest-known radio pulsars for example,
it is not hard to realize that these objects are compact enough that they should really be treated 
in a general relativistic manner. To be more concrete, note that the values of the parameters
characterizing typical radio pulsars are $r_{0}({\rm radius}) \sim 10^{6} (cm)$, 
$M({\rm mass}) \sim 1.4 M_{\odot} \sim 2\times 10^{33} (g)$, 
$\tau({\rm pulsation ~period}) \sim 10^{-3} - 1 (sec)$, and
$B({\rm magnetic ~field ~strength}) \sim 10^{12} (G)$. 
Thus, the Schwarzschild radius (gravitational radius) of a typical radio pulsar is estimated to
be $r_{Sch} = 2GM_{\odot}/c^2 \sim 3\times 10^{5} (cm)$ (where $G$ and $c$ denote Newton's
constant and the speed of light respectively) ; hence, one ends up with the ratio
$r_{0}/r_{Sch} \sim 10^{6} (cm)/3\times 10^{5} (cm) \sim 3$. This simple argument indicates that, indeed,
even the radio pulsar (which is perhaps the least energetic among all of pulsars) is a
highly self-gravitating compact object that needs to be treated relativistically. Of course,
a ``relativistic'' treatment here means that the region surrounding the pulsar, namely, the
pulsar magnetosphere, has to be described by using a curved spacetime rather than simply a flat
one. Then, the question now boils down to this ; what would be the relevant metric for describing
the vicinity of a rotating neutron star ?  Although it does not seem to be well-known, fortunately,
we have such a metric for the region exterior to {\it slowly-rotating} relativistic stars, such as
neutron stars, white dwarfs, and supermassive stars, and it is the one constructed long ago by
Hartle and Thorne \cite{ht}. Thus, the Hartle-Thorne metric is a stationary axisymmetric solution to
the vacuum Einstein equation and in the present work, we shall take this Hartle-Thorne metric as 
a relavant one to represent the spacetime exterior to slowly-rotating neutron stars.  

As we mentioned above, here in the present work, we are particularly interested in the exact 
theoretical determination of the amount of charge on and the structure of the magnetic field around 
a slowly-rotating neutron star by treating its exterior region literally as a curved spacetime. 
As such, it seems relevant to introduce a related recent work which, as a matter of fact, has been
one of the motivations for this study.  In our earlier work \cite{hongsu}, we explored how 
much of the magnetic flux could actually
penetrate the horizon of a rotating black hole in association with the Blandford-Znajek mechanism \cite{bz}
for rotational energy extraction from a Kerr hole. In doing so, we made use of the algorithm
suggested by Wald \cite{wald} to figure out the amount of charge on and the structure of the magnetic field around
a rotating black hole. The results presented there for the case of a rotating
black hole can, therefore, be compared with those that will be presented in this work for the case 
of a rotating neutron star because, as stated, we shall in this work employ the algorithm of Wald to 
determine the amount of charge accretion on and the structure of the magnetic field around a rotating 
neutron star, as well. Thus, it should be interesting to read the present work in 
parallel with our earlier work \cite{hongsu} on Kerr black hole electrodynamics. 

Speaking of the Blandford-Znajek mechanism, perhaps it would be relevant to briefly sketch their
idea here as it has a significant impact on the physics underlying both the black hole and the 
(rotating) neutron star's magnetosphere. According to the Blandford-Znajek mechanism \cite{bz},
the way some of the hole's rotational energy is carried off by the magnetic field lines 
threading the hole can be summarized as follows : Angular momentum and rotational 
energy transport is
achieved in terms of the conservation of the ``electromagnetic'' angular momentum and the energy flux flowing 
from the hole to its nearby field particularly when the force-free condition is satisfied in the 
surrounding magnetosphere. It is also amusing to note that the Blandford-Znajek mechanism can be 
translated into a simple-minded circuit analysis. In this alternative picture, the current flowing 
from near the pole toward the equator of the rotating black hole geometry and the
poloidal magnetic field together generate a ``magnetic braking'' torque that is antiparallel to the 
hole's spin direction. As a result, the rotating hole, which is a part of the circuit, spins down, 
and its angular momentum and the rotational energy are extracted. 
For more details, we refer the reader to the original work of Blandford and Znajek \cite{bz} and 
some comprehensive review articles \cite{hklee} on this topic, including the ``membrane paradigm'' 
approach \cite{membrane}. 

\begin{center}
{\rm\bf II. THE HARTLE-THORNE METRIC FOR THE REGION EXTERIOR TO SLOWLY-ROTATING NEUTRON STARS}
\end{center}

The Hartle-Thorne metric \cite{ht} is given by
\begin{eqnarray}
ds^2 = -\alpha^2 dt^2 + g_{rr}dr^2 + g_{\theta\theta}d\theta^2 + g_{\phi\phi}
(d\phi + \beta^{\phi}dt)^2, 
\end{eqnarray}
where
\begin{eqnarray}
\alpha^2 &=& \Delta R, ~~~\beta^{\phi} = -\omega = -{2J\over r^3}, \nonumber \\
g_{rr} &=& {S\over \Delta}, ~~~g_{\theta\theta} = r^2 A, 
~~~g_{\phi\phi} = \varpi^2 = r^2 A\sin^2 \theta, \\
g_{tt} &=& -[\alpha^2 - (\beta^{\phi})^2g_{\phi\phi}] = - [\Delta R - {4J^2\over r^4}A\sin^2 \theta],
\nonumber \\
g_{t\phi} &=& \beta^{\phi} g_{\phi\phi} = -{2J\over r}A\sin^2 \theta, \nonumber
\end{eqnarray}
and
\begin{eqnarray}
\Delta &=& \left(1 - {2M\over r} + {2J^2\over r^4}\right), \nonumber \\
R &=& \left[1 + 2\left\{{J^2\over Mr^3}\left(1+{M\over r}\right) + {5\over 8}
{Q-J^2/M \over M^3}Q^2_{2}({r\over M}-1)\right\}P_{2}(\cos \theta)\right], \\
S &=& \left[1 - 2\left\{{J^2\over Mr^3}\left(1-{5M\over r}\right) + {5\over 8}
{Q-J^2/M \over M^3}Q^2_{2}({r\over M}-1)\right\}P_{2}(\cos \theta)\right], \nonumber \\
A &=& 1 + 2\left[-{J^2\over Mr^3}\left(1+{2M\over r}\right) \right. \nonumber \\
&&\left. + {5\over 8}{Q-J^2/M \over M^3}\left\{{2M\over [r^2(1-2M/r)]^{1/2}}Q^1_{2}({r\over M}-1)
- Q^2_{2}({r\over M}-1)\right\}\right]P_{2}(\cos \theta) \nonumber
\end{eqnarray}
with $M$, $J$ and $Q$ being the mass, the angular momentum, and the
mass quadrupole moment of the slowly rotating neutron star, respectively,
$P_{2}(\cos \theta) = (3\cos^2 \theta -1)/2$ being the Legendre polynomial, and 
$Q^{m}_{n}$ being the associated Legendre polynomial, namely,
\begin{eqnarray}
Q^{1}_{2}(z) &=& (z^2-1)^{1/2}\left[{3z^2 - 2\over z^2 - 1} 
- {3\over 2}z\log \left({z+1\over z-1}\right)\right], \\
Q^{2}_{2} (z) &=& \left[{3\over 2}(z^2 - 1)\log \left({z+1\over z-1}\right) 
- {3z^3 - 5z\over z^2 - 1}\right] ; \nonumber
\end{eqnarray}
hence,
\small
\begin{eqnarray}
Q^{1}_{2}({r\over M}-1) &=& {r\over M}\left(1-{2M\over r}\right)^{1/2}
\left[{3(r/M)^2(1-2M/r)+1 \over (r/M)^2(1-2M/r)}
+ {3\over 2}{r\over M}\left(1-{M\over r}\right)\log \left(1-{2M\over r}\right)\right], \\
Q^{2}_{2}({r\over M}-1) &=& -\left[{3\over 2}\left({r\over M}\right)^2\left(1-{2M\over r}\right)
\log \left(1-{2M\over r}\right) + 
{{M/r}(1-M/r)\left\{3(r/M)^2(1-2M/r)-2\right\} \over 
(1-2M/r)}\right]. \nonumber
\end{eqnarray}
\normalsize
As is well-known, the only known exact metric solution exterior to a rotating object is the
Kerr metric \cite{kerr}. Thus, it would be worth clarifying the relation of the Hartle-Thorne metric for
slowly-rotating relativistic stars given above to the Kerr metric. As Hartle and Thorne \cite{ht},  
take the Kerr metric given in Boyer-Lindquist coordinate and expand it to second order
in angular velocity (namely, the angular shift $\beta^{\phi}$) followed by a coordinate
transformation in the $(r, \theta)$-sector,
\begin{eqnarray}
&&r \to  r\left[1-{a^2\over 2r^2}\left\{(1+{2M\over r})(1-{M\over r})+\cos^2 \theta
(1-{2M\over r})(1+{3M\over r})\right\}\right], \nonumber \\
&&\theta \to  \theta - a^2 \cos \theta \sin \theta {1\over 2r^2}(1+{2M\over r}), 
\end{eqnarray}
where $a=J/M$.
Then, one can realize that the resulting expanded Kerr metric coincides with the particular case
$Q=J^2/M$ (with $Q$ being the mass quadrupole moment of the rotating object) of the Hartle-Thorne
metric. Therefore, in general, this Hartle-Thorne metric is {\it not} a slow-rotation limit of the Kerr 
metric. Rather, the slow-rotation limit of the Kerr metric is a special case of this more general 
Hartle-Thorne metric. As a result, the Hartle-Thorne metric with an arbitrary value of the mass
quadrupole moment $Q$ can generally describe a slowly-rotating neutron star of any shape
(as long as it retains the axisymmetry). 

Next, since we shall employ in the present work the Hartle-Thorne metric to represent the
spacetime exterior to slowly-rotating neutron stars, we would like to carefully distinguish
between the horizon radius of the Hartle-Thorne metric and the actual radius of the neutron
stars. Indeed, one of the obvious differences between the black hole case and the neutron star
case is the fact that the black hole is characterized by its event horizon while the neutron
star has a hard surface. Since this solid surface of a neutron star (which we shall henceforth denote
by $r_{0}$) lies outside of its gravitational radius, which amounts to the Killing horizon radius 
of the Hartle-Thorne metric, $r_{H}$ at which $\Delta = (1 - 2M/r + 2J^2/r^4) = 0$ in Eq. (3),
we have $r_{0}>r_{H}$. In the present work, however, we shall never speak of the Killing horizon of 
the Hartle-Thorne metric as it is an irrelevant quantity playing no physical role. 

Now, we turn to the choice of an orthonormal tetrad frame and we shall particularly choose the
zero-angular-momentum-observer (ZAMO) \cite{zamo} frame, which is a sort of fiducial observer (FIDO) frame.
Generally speaking, in order to represent a given background geometry, one needs to first 
choose the coordinate system in which the metric is to be given ; next, in order to obtain 
the physical components of a tensor (such as the electric and the magnetic field values), one has to 
select a tetrad frame (in a given coordinate system) to which the tensor components are to be
projected. As is well-known, the orthonormal tetrad is a set of four mutually orthogonal unit 
vectors at each point in a given spacetime. And these unit vectors give the directions of the four axes of 
a locally-Minkowskian coordinate system. Such an orthonormal tetrad associated with the 
Hartle-Thorne metric given above may be chosen as  
$e^{A}= e^{A}_{\mu}dx^{\mu} = (e^{0}, e^{1}, e^{2}, e^{3})$,
\ba
e^{0} &=& \alpha dt = (\Delta R)^{1/2}dt, \nonumber  \\
e^{1} &=& g^{1/2}_{rr}dr = \left({S\over \Delta}\right)^{1/2}dr, \\
e^{2} &=& g^{1/2}_{\theta \theta}d\theta = rA^{1/2}d\theta, \nonumber  \\
e^{3} &=& g^{1/2}_{\phi \phi}(d\phi + \beta^{\phi}dt) =
r\sin \theta A^{1/2}\left[d\phi - {2J\over r^3}dt\right], \nonumber
\ea
and its dual basis is given by
$e_{A}= e^{\mu}_{A}\partial_{\mu} = 
(e_{0}=e_{(t)}, e_{1}=e_{(r)}, e_{2}=e_{(\theta)}, e_{3}=e_{(\phi)})$,
\ba
e_{0} &=& {1\over \alpha}(\partial_{t}-\beta^{\phi} \partial_{\phi}) = 
(\Delta R)^{-1/2}\left[\partial_{t}+\frac{2J}{r^3}\partial_{\phi}\right], \nonumber  \\
e_{1} &=& g^{-1/2}_{rr}\partial_{r} = \left(\frac{\Delta}{S}\right)^{1/2}\partial_{r}, \\
e_{2} &=& g^{-1/2}_{\theta \theta}\partial_{\theta} = 
\frac{1}{rA^{1/2}}\partial_{\theta}, \nonumber  \\
e_{3} &=& g^{-1/2}_{\phi \phi}\partial_{\phi} = 
\frac{1}{rA^{1/2}\sin \theta}\partial_{\phi}. \nonumber
\ea

A local, stationary observer at rest in this orthonormal tetrad frame $e^{A}$ has a worldline
given by $\{dr=0, ~d\theta=0, ~(d\phi + \beta^{\phi}dt)=0\}$, which is orthogonal to spacelike 
hypersurfaces and has orbital angular velocity given by
\be
\omega = \frac{d\phi}{dt} = -\beta^{\phi} = -\frac{g_{t\phi}}{g_{\phi \phi}}
= \frac{2J}{r^3}.
\ee
This is the long-known {\it Lense-Thirring} precession \cite{lt} angular velocity arising due to
the ``dragging of inertial frame'' effect of a stationary axisymmetric spacetime. Indeed, it is
straightforward to demonstrate that this orthonormal tetrad observer can be identified with a
ZAMO carrying zero intrinsic angular momentum. 
To this end, recall that when a spacetime metric
possesses a rotational (azimuthal) isometry, there exists an associated rotational Killing field
$\psi^{\mu} = (\partial /\partial \phi)^{\mu} = \delta^{\mu}_{\phi}$ such that its inner product
with the tangent (velocity) vector $u^{\mu} = dx^{\mu}/d\tau$ (with $\tau $ denoting the
particle's proper time) of the geodesic of a test particle is constant along the geodesic, i.e.,
\ba
\tilde{L} &=& g_{\alpha \beta}\psi^{\alpha}u^{\beta} = 
g_{\phi t}\psi^{\phi}u^{t} + g_{\phi \phi}\psi^{\phi}u^{\phi} \nonumber \\
&=& g_{\phi t}\frac{dt}{d\tau} + g_{\phi \phi}\frac{d\phi}{d\tau}.
\ea
Now, particularly when a local, stationary observer, which here is taken to be a test particle, 
carries zero angular momentum, $\tilde{L} = 0$, the angular velocity becomes
\be
\omega = \frac{d\phi}{dt} = \frac{(d\phi/d\tau)}{(dt/d\tau)} = 
-\frac{g_{t\phi}}{g_{\phi \phi}} = -\beta^{\phi},
\ee
which confirms the identification of the local observer at rest in this orthonormal tetrad
frame given above as a ZAMO.

\begin{center}
{\rm\bf III. SOLUTION-GENERATING METHOD BY WALD}
\end{center}

{\bf 1. Wald Field}

From the general properties of Killing fields \cite{papa}, {\it a Killing vector in
a vacuum spacetime generates a solution of Maxwell's equations in the
background of that vacuum spacetime}, long ago, Wald \cite{wald} constructed a stationary
axisymmetric solution of Maxwell's equations in Kerr black hole spacetime. 
To be a little more concrete, Wald's construction is based on the following two
theorems : \\
(A) {\it The axial Killing vector $\psi^{\mu} = (\partial/\partial \phi)^{\mu}$
generates a stationary axisymmetric test electromagnetic field which asymptotically
approaches a uniform magnetic field, has no magnetic monopole moment and has 
charge $= 4J$:} \\
$F_{\psi} = d\psi $ \\
where ``$d$'' denotes the exterior derivative and $J$ is the angular momentum of a 
Kerr black hole. \\
(B) {\it The time translational Killing vector $\xi^{\mu} = (\partial/\partial t)^{\mu}$
generates a stationary axisymmetric test electromagnetic field which vanishes
asymptotically, has no magnetic monopole moment and has
charge $= -2M$:} \\
$F_{\xi} = d\xi $ \\
where $M$ is the mass of the Kerr hole. 

Apparently, this solution generating method of Wald can be applied in a straightforward manner
to the construction of a stationary axisymmetric solution of Maxwell's equations in a spacetime
surrounding a slowly-rotating neutron star represented by the Hartle-Thorne metric given earlier.
Thus, we now look for the solution of the electromagnetic test field $F$ that occurs when 
a stationary axisymmetric slowly-rotating neutron star of mass $M$ and angular momentum $J$
(represented by the Hartle-Thorne metric) 
is placed in an originally uniform magnetic field of strength $B_{0}$ aligned along the symmetry 
axis of the neutron star. For now, we consider the case when the electric charge is absent. 
Obviously, then the solution can be readily written down, by referring to theorems (A) and (B) 
above, as follows :
\begin{eqnarray}
F = {1\over 2}B_{0} \left[F_{\psi}-\left({4J\over -2M}\right)F_{\xi}\right]
= {1\over 2}B_{0} \left[d\psi + {2J\over M} d\xi \right].
\end{eqnarray}
Then, using  $F={1\over 2}F_{\mu\nu}dx^{\mu}\wedge dx^{\nu}$ and 
$\psi = \psi_{\mu}dx^{\mu}$, $\xi = \xi_{\nu}dx^{\nu}$, with
\begin{eqnarray}
\xi_{\mu} &=& g_{\mu\nu}\xi^{\nu}=g_{\mu\nu}\delta^{\nu}_{t}=g_{\mu t},\\
\psi_{\mu} &=& g_{\mu\nu}\psi^{\nu}=g_{\mu\nu}\delta^{\nu}_{\phi}=g_{\mu\phi}
\nonumber
\end{eqnarray}
and for the Hartle-Thorne metric given in the previous subsection, we can write the solution
above in a concrete form as 
\begin{eqnarray}
F &=& \frac{1}{2}B_{0}\left[(\partial_{r}g_{t\phi})+\frac{2J}{M}(\partial_{r}g_{tt})\right]
(dr \wedge dt) 
  + \frac{1}{2}B_{0}\left[(\partial_{\theta}g_{t\phi})+\frac{2J}{M}(\partial_{\theta}g_{tt})\right]
(d\theta \wedge dt) \nonumber \\
  &+& \frac{1}{2}B_{0}\left[(\partial_{r}g_{\phi \phi})+\frac{2J}{M}(\partial_{r}g_{\phi t})\right]
(dr \wedge d\phi) 
  + \frac{1}{2}B_{0}\left[(\partial_{\theta}g_{\phi \phi})+\frac{2J}{M}(\partial_{\theta}g_{\phi t})\right]
(d\theta \wedge d\phi)]. 
\end{eqnarray}
Here
\small
\begin{eqnarray}
(\partial_{r}g_{tt}) &=& 
- \left(\frac{2M}{r^2}-\frac{8J^2}{r^5}\right)R + 2\Delta \left[\frac{J^2}{Mr^4}
\left(3+\frac{4M}{r}\right) - \frac{5}{8} \frac{Q-J^2/M}{M^3}Q^{'2}_{2}(\frac{r}{M}-1)\right]P_{2}(\cos \theta) \nonumber \\
&-& \frac{16J^2}{r^5}A\sin^2 \theta
+ \frac{8J^2}{r^4}\sin^2 \theta \left[\frac{J^2}{Mr^4}\left(3+\frac{8M}{r}\right) + \frac{5}{8} \frac{Q-J^2/M}{M^3}\times \right. \\
&&\left. \left\{\frac{-2M(1-M/r)}{r^2 (1-2M/r)^{3/2}}Q^{1}_{2}(\frac{r}{M}-1)
+ \frac{2M}{[r^2 (1-2M/r)]^{1/2}}Q^{'1}_{2}(\frac{r}{M}-1) - Q^{'2}_{2}(\frac{r}{M}-1)\right\}\right]P_{2}(\cos \theta), \nonumber \\
(\partial_{\theta}g_{tt}) &=& 
6\Delta \left[\frac{J^2}{Mr^3}
\left(1+\frac{M}{r}\right) + \frac{5}{8} \frac{Q-J^2/M}{M^3}Q^{2}_{2}(\frac{r}{M}-1)\right]\sin \theta \cos \theta \nonumber \\
&+& \frac{8J^2}{r^4}A\sin \theta \cos \theta
- \frac{24J^2}{r^4}\left[-\frac{J^2}{Mr^3}\left(1+\frac{2M}{r}\right)  \right. \\
&&\left. + \frac{5}{8} \frac{Q-J^2/M}{M^3}\left\{
+ \frac{2M}{[r^2 (1-2M/r)]^{1/2}}Q^{1}_{2}(\frac{r}{M}-1) - Q^{2}_{2}(\frac{r}{M}-1)\right\}\right]\sin^3 \theta \cos \theta, \nonumber \\
(\partial_{r}g_{t\phi}) &=& 
\frac{2J}{r^2}A\sin^2 \theta
- \frac{4J}{r}\sin^2 \theta \left[\frac{J^2}{Mr^4}\left(3+\frac{8M}{r}\right) + \frac{5}{8} \frac{Q-J^2/M}{M^3}\times \right. \\
&&\left. \left\{\frac{-2M(1-M/r)}{r^2 (1-2M/r)^{3/2}}Q^{1}_{2}(\frac{r}{M}-1)
+ \frac{2M}{[r^2 (1-2M/r)]^{1/2}}Q^{'1}_{2}(\frac{r}{M}-1) - Q^{'2}_{2}(\frac{r}{M}-1)\right\}\right]P_{2}(\cos \theta), \nonumber \\
(\partial_{\theta}g_{t\phi}) &=& 
- \frac{4J}{r}A\sin \theta \cos \theta
+ \frac{12J}{r}\left[-\frac{J^2}{Mr^3}\left(1+\frac{2M}{r}\right) \right. \\
&&\left. + \frac{5}{8} \frac{Q-J^2/M}{M^3}\left\{
+ \frac{2M}{[r^2 (1-2M/r)]^{1/2}}Q^{1}_{2}(\frac{r}{M}-1) - Q^{2}_{2}(\frac{r}{M}-1)\right\}\right]\sin^3 \theta \cos \theta, \nonumber \\
(\partial_{r}g_{\phi \phi}) &=& 
2rA\sin^2 \theta
+ 2r^2 \sin^2 \theta \left[\frac{J^2}{Mr^4}\left(3+\frac{8M}{r}\right) + \frac{5}{8} \frac{Q-J^2/M}{M^3}\times \right. \\
&&\left. \left\{\frac{-2M(1-M/r)}{r^2 (1-2M/r)^{3/2}}Q^{1}_{2}(\frac{r}{M}-1)
+ \frac{2M}{[r^2 (1-2M/r)]^{1/2}}Q^{'1}_{2}(\frac{r}{M}-1) - Q^{'2}_{2}(\frac{r}{M}-1)\right\}\right]P_{2}(\cos \theta), \nonumber \\
(\partial_{\theta}g_{\phi \phi}) &=& 
2r^2 A\sin \theta \cos \theta
- 6r^2 \left[-\frac{J^2}{Mr^3}\left(1+\frac{2M}{r}\right) \right. \\
&&\left. + \frac{5}{8} \frac{Q-J^2/M}{M^3}\left\{
+ \frac{2M}{[r^2 (1-2M/r)]^{1/2}}Q^{1}_{2}(\frac{r}{M}-1) - Q^{2}_{2}(\frac{r}{M}-1)\right\}\right]\sin^3 \theta \cos \theta, \nonumber 
\end{eqnarray}
\normalsize
where
\small
\begin{eqnarray}
&&Q^{'1}_{2}(\frac{r}{M}-1) = \frac{d}{dr}Q^{1}_{2}(\frac{r}{M}-1)  \\
&&={1\over r}\left(1-{2M\over r}\right)^{-1/2}
\left[{{M/r}(1-M/r)\left\{6(r/M)^2(1-2M/r)-1\right\} \over (1-2M/r)} \right. \nonumber \\
&&\left. + {3\over 2}\left\{2\left({r\over M}\right)^2\left(1-{2M\over r}\right)+1\right\}\log \left(1-{2M\over r}\right)\right], \nonumber \\
&&Q^{'2}_{2}({r\over M}-1) = \frac{d}{dr}Q^{2}_{2}(\frac{r}{M}-1) \\
&&= -\frac{1}{M}\left[6 + 3\left({r\over M}\right)\left(1-{M\over r}\right)
\log \left(1-{2M\over r}\right) + 
2\left({M\over r}\right)^2(1-{2M\over r})^{-1} + 4\left({M\over r}\right)^4(1-{2M\over r})^{-2}\right]. \nonumber
\end{eqnarray}
\normalsize
Thus, this can be thought of as the Wald-type field for the case of a rotating neutron star.
Of course, caution needs to be exercised, the point being that this Maxwell field is correct
only for a slowly-rotating neutron star, but not for a neutron star with arbitrary rotation. 

{\bf 2. Wald Charge}

We now turn to the issue of {\it charge accretion} onto a slowly-rotating neutron star immersed in a 
magnetic field surrounded by an ionized interstellar medium (``plasma''). We shall essentially
follow the argument again given by Wald \cite{wald}, and to do so, we first need to know the physical components
of the electric and the magnetic fields. This can be achieved by projecting the Maxwell field tensor
given above in Eq. (14) onto the ZAMO tetrad frame. Thus, using the dual to the ZAMO tetrad,
$e_{A}=(e_{0}=e_{(t)}, e_{1}=e_{(r)}, e_{2}=e_{(\theta)}, e_{3}=e_{(\phi)})$, 
given earlier in Eq. 8), we can now read off the ZAMO tetrad components of $F_{\mu \nu}$ as
\ba
F_{10} &=& F_{\mu\nu} e^{\mu}_{1}e^{\nu}_{0} \nonumber \\
      &=& \left(\frac{\Delta}{S}\right)^{1/2}\left[(\Delta R)^{-1/2}\frac{1}{2}B_{0}
\left\{(\partial_{r}g_{t\phi})+\frac{2J}{M}(\partial_{r}g_{tt})\right\} 
 +(\Delta R)^{-1/2}\frac{2J}{r^3}\frac{1}{2}B_{0}
\left\{(\partial_{r}g_{\phi \phi})+\frac{2J}{M}(\partial_{r}g_{\phi t})\right\}\right], \nonumber \\ 
F_{20} &=& F_{\mu\nu} e^{\mu}_{2}e^{\nu}_{0} \nonumber \\
      &=&  \frac{1}{rA^{1/2}}\left[(\Delta R)^{-1/2}\frac{1}{2}B_{0}
\left\{(\partial_{\theta}g_{t\phi})+\frac{2J}{M}(\partial_{\theta}g_{tt})\right\} 
 +(\Delta R)^{-1/2}\frac{2J}{r^3}\frac{1}{2}B_{0}
\left\{(\partial_{\theta}g_{\phi \phi})+\frac{2J}{M}(\partial_{\theta}g_{\phi t})\right\}\right], \nonumber \\
F_{30} &=& F_{\mu\nu} e^{\mu}_{3}e^{\nu}_{0} = 0, \\
F_{12} &=& F_{\mu\nu} e^{\mu}_{1}e^{\nu}_{2} = 0, \nonumber \\
F_{13} &=& F_{\mu\nu} e^{\mu}_{1}e^{\nu}_{3} \nonumber \\
      &=&  \left(\frac{\Delta}{S}\right)^{1/2} \frac{1}{rA^{1/2}\sin \theta}
\frac{1}{2}B_{0}\left\{(\partial_{r}g_{\phi \phi})+\frac{2J}{M}(\partial_{r}g_{\phi t})\right\}, \nonumber \\
F_{23} &=& F_{\mu\nu} e^{\mu}_{2}e^{\nu}_{3} \nonumber \\
      &=& \frac{1}{rA^{1/2}}\frac{1}{rA^{1/2}\sin \theta}
\frac{1}{2}B_{0}\left\{(\partial_{\theta}g_{\phi \phi})+\frac{2J}{M}(\partial_{\theta}g_{\phi t})\right\},
            \nonumber                          
\ea
where $(\partial_{r}g_{t\phi})$, $(\partial_{r}g_{tt})$, $(\partial_{\theta}g_{t\phi})$,
$(\partial_{\theta}g_{tt})$, $(\partial_{r}g_{\phi \phi})$, and
$(\partial_{\theta}g_{\phi \phi})$ are as given above in Eqs. (15)-(20).  

Here, consider particularly the radial component of the electric field (as observed by a local
observer in this ZAMO tetrad frame) which, along the symmetry axis ($\theta = 0, ~\pi$) of the 
slowly-rotating neutron star, becomes 
\begin{eqnarray}
&&E_{\hat{r}} = E_{1} = F_{10} \\
&&= \frac{1}{2}B_{0}(RS)^{-1/2}\frac{2J}{M}\left[-\left(\frac{2M}{r^2}-\frac{8J^2}{r^5}\right)R
+ 2\Delta \left\{\frac{J^2}{Mr^4}\left(3+\frac{4M}{r}\right) - \frac{5}{8}\frac{Q-J^2/M}{M^3}
Q^{'2}_{2}(\frac{r}{M}-1)\right\}\right]. \nonumber
\end{eqnarray}
Note that from this expression for the radial component of the electric field, it is {\it not}
possible to determine the sign of charge accreted in terms of the relative direction between the angular
momentum of the neutron star and the external magnetic field. This point is in contrast to
what happens in the case of the Kerr black hole\cite{wald, hongsu} studied previously.
To be a little more concrete, it was realized that the radial component of the
electric field for the case of Kerr black hole is radially {\it inward/outward}
if the hole's axis of rotation and the external magnetic field are {\it parallel/antiparallel}.
Put differently, this implies that if the spin of the hole and the magnetic field are {\it parallel},
then {\it positively charged} particles on the symmetry axis of the hole will be
pulled into the hole while if the spin of the hole and the magnetic field are {\it antiparallel},
{\it negatively charged} particles on the symmetry axis of the hole will be
pulled into the hole. In this manner, a rotating black hole will ``selectively'' accrete charged
particles until it builds up an ``equilibrium'' net charge. For the present case of a rotating neutron
star, however, an argument of this sort cannot be presented due essentially to the added complexity 
of the Hartle-Thorne metric compared to the Kerr metric. Nevertheless, one can still ask and answer
the next natural question ; how then can the equilibrium net charge be determined ?  
To answer this question, we resort to the ``injection energy'' argument originally proposed by 
Carter \cite{carter} for the Kerr black hole case. Again, it can be applied equally well to the present case
of a slowly-rotating neutron star. 

Recall first that the energy of a charged 
particle in a stationary spacetime with time translational isometry generated by the Killing 
field $\xi^{\mu}=(\partial /\partial t)^{\mu}$ in the presence of a stationary electromagnetic 
field is given by 
\begin{eqnarray}
\varepsilon = - p_{\alpha}\xi^{\alpha} = - g_{\alpha\beta}p^{\alpha}\xi^{\beta} \nonumber
\end{eqnarray}
with $p^{\mu}={\tilde m}u^{\mu}-eA^{\mu}$ being the 4-momentum of a charged particle with mass
and charge $\tilde{m}$ and $e$, respectively. Now, if we lower the charged particle down the
symmetry axis into the rotating neutron star, the change in electrostatic energy of the particle will be
\begin{eqnarray}
\delta \varepsilon = \varepsilon_{final} - \varepsilon_{initial} =
eA_{\alpha}\xi^{\alpha}|_{horizon} - eA_{\alpha}\xi^{\alpha}|_{\infty}.
\end{eqnarray}
Now, if $\delta \varepsilon < 0$, it will be energetically favorable for the neutron star
to accrete particles with this charge whereas if $\delta \varepsilon > 0$,  
it will accrete particles with opposite charge. In either case, the rotating neutron star will
selectively accrete charge until $A^{\mu}$ is changed sufficiently that the electrostatic ``injection
energy'' $\delta \varepsilon$ is reduced to zero. We are, then, ready to determine, by this injection
energy argument due to Carter, what the equilibrium net charge accreted onto the neutron star would be.

In the discussion of the Wald-type field given above, we only restricted ourselves to the case of solutions to
Maxwell equation in the background of an {\it uncharged} stationary axisymmetric neutron star spacetime,
and it was given by Eq. (12). Now, we need the solution when the stationary axisymmetric neutron star is
slightly charged via the charge accretion process described above. Then, according to {\it theorem
(B)} in the discussion of the Wald field given earlier, there can be, at most, one more perturbation of a
stationary axisymmetric vacuum (neutron star) spacetime which corresponds to adding an electric 
charge $\tilde{Q}$ to the neutron star, which is nothing more than a linearly superposition of the solution 
$(-\tilde{Q}/2M)F_{\xi} = (-\tilde{Q}/2M)d\xi$ to the solution given in Eq. (12) to get 
\begin{eqnarray}
F = \frac{1}{2}B_{0}[d\psi + \frac{2J}{M}d\xi]-\frac{\tilde{Q}}{2M}d\xi,
\end{eqnarray} 
which, in terms of the gauge potental, amounts to
\begin{eqnarray}
A_{\mu} = {1\over 2}B_{0}(\psi_{\mu}+{2J\over M}\xi_{\mu})-{\tilde{Q}\over 2M}\xi_{\mu}.
\end{eqnarray}
Then, the electrostatic injection energy can be computed as
\begin{eqnarray}
\delta \varepsilon &=& eA_{\alpha}\xi^{\alpha}|_{horizon} - eA_{\alpha}\xi^{\alpha}|_{\infty}
\nonumber \\
&=& e\left(\frac{B_{0}J}{M}-\frac{\tilde{Q}}{2M}\right)[1 - \Delta R|_{r=r_{0}, \theta =0, \pi}],
\end{eqnarray}
where we used $\xi_{\mu}\xi^{\mu} = -\Delta R|_{r=r_{0}, \theta =0, \pi}$ 
(at $r=r_{0}, \theta =0, \pi $),  $\xi_{\mu}\xi^{\mu} = -1$ (at $r\to \infty, \theta =0, \pi $),
and $\psi_{\mu}\xi^{\mu} = 0$ (on the symmetry axis $\theta =0, \pi $) and $r_{0}$ denotes the 
radius of the neutron star. As was discussed carefully in Section II, $r_{0}$ is located
outside of the gravitational radius of the rotating neutron star, which amounts, in our description, 
to the Killing horizon radius of the Hartle-Thorne metric at which $\Delta = 0$.
Thus, one may conclude that a rotating neutron star in a uniform magnetic field will accrete charge until
the gauge potential evolves to a value at which $\delta \varepsilon = 0$, yielding an equilibrium
net charge of $\tilde{Q} = 2B_{0}J$. Thus, this amount of charge may be called the Wald-type charge, and it is
particularly interesting to note that it turns out to be the same as the ``Wald charge'' (the 
equilibrium net charge) for the case of a Kerr black hole \cite{wald,  hongsu}. This result appears
to imply that the amount of equilibrium net charge $\tilde{Q} = 2B_{0}J$ might be a generic value once a
general relativistic rotating object gets charged via accretion. 

There, however, does exist a
distinction between the case of a Kerr black hole and that of a rotating neutron star.
That is, from Eq. (28), the rotating neutron star will eventually accrete an equilibrium net charge
of $\tilde{Q} = 2B_{0}J$ only if $[1 - \Delta R|_{r=r_{0}, \theta =0, \pi}] \neq 0$. Certainly, this feature
was absent in the case of a Kerr black hole and can be attributed to the different structure of the
Hartle-Thorne metric describing the region surrounding the rotating neutron star.
Therefore, particularly if the values of mass ($M$), angular momentum ($J$), mass quadrupole 
moment ($Q$), and radius along the symmetry axis ($r_{0}$) are such that they satisfy
\begin{eqnarray}
&&\Delta R|_{r=r_{0}, \theta =0, \pi} \\
&&= \left(1 - {2M\over r_{0}} + {2J^2\over r^4_{0}}\right)
\left[1 + 2\left\{{J^2\over Mr^3_{0}}\left(1+{M\over r_{0}}\right) + {5\over 8}
{Q-J^2/M \over M^3}Q^2_{2}({r_{0}\over M}-1)\right\}\right] = 1 \nonumber
\end{eqnarray}
then it is ``energetically unlikely'' that a slowly-rotating neutron star would accrete
any charge, and we find a peculiar feature which has no parallel in the Kerr black hole case. 
It, however, should be taken with some caution (i.e., should not be taken as being conclusive) because
the injection energy argument due to Carter that leads to this conclusion is, rigorously speaking,
restricted to the situation along the symmetry axis $\theta = 0, \pi$, and at this point it
is not so obvious that the same would be true off the symmetry axis. 

Next, recall that we announced from the beginning that we shall consider the case when the charge
accreted on the neutron star is small enough not to distort the background Hartle-Thorne geometry. Now,
we provide the rationale that this is, indeed, what can actually happen. To do so, we first
assume that just as in the case of a Kerr black hole (a particular slow-rotation limit of which is
the Hartle-Thorne geometry), in the present Hartle-Thorne spacetime, $J\leq M^2$. 
Then, since the typical value of the charge accreted on
the rotating neutron star is $\tilde{Q}=2B_{0}J$, as just described, its charge-to-mass
ratio has an upper bound
\begin{eqnarray}
{\tilde{Q} \over M} = 2B_{0}\left({J\over M}\right) \leq 2B_{0}M = 2\left({B_{0}\over 10^{15}(G)}\right)
\left({M\over M_{\odot}}\right)10^{-5}, \nonumber
\end{eqnarray}
where in the last equality we have converted $B_{0}$ and $M$ from geometrized units to
solar-mass units and gauss \cite{hongsu}. Thus, for the typical case of a radio pulsar with mass 
$M\sim ~M_{\odot}$ in a surrounding magnetic field of strength $B_{0}\sim 10^{12} (G)$, 
$\tilde{Q}/M \sim 10^{-8} << 1$.  Thus, as we can see in this radio pulsar case,
the charge-to-mass ratio of the associated rotating neutron star is small enough not to disturb 
the geometry itself. Thus, we can safely employ the solution-generating method suggested by Wald 
to construct the solution to Maxwell's equations when some amount of charge is around, 
to which we now turn.  

\begin{center}
{\rm\bf IV. STATIONARY AXISYMMETRIC MAXWELL FIELD AROUND A ``SLIGHTLY CHARGED'' AND SLOWLY-ROTATING NEUTRON STAR}
\end{center}

As we discussed in the previous subsection, the stationary
axisymmetric solution to Maxwell's equation in the background of a rotating neutron star with charge $\tilde{Q}$
accreted in an originally uniform magnetic field can be constructed as    
\ba
F =\frac{1}{2}B_{0}[F_{\psi}-(\frac{4J}{-2M})F_{\xi}]+(\frac{\tilde{Q}}{-2M})F_{\xi}
=\frac{1}{2}B_{0}[d\psi + \frac{2J}{M}(1-\frac{\tilde{Q}}{2B_{0}J})d\xi]. 
\ea
Again for the Hartle-Thorne metric for a slowly-rotating neutron star, which was given in the previous 
subsection, the solution above can be written as 
\ba
F &=& \frac{1}{2}B_{0}\left[(\partial_{r}g_{t\phi})+
\frac{2J}{M}\left(1 - \frac{\tilde{Q}}{2B_{0}J}\right)(\partial_{r}g_{tt})\right]
(dr \wedge dt) \nonumber \\
  &+& \frac{1}{2}B_{0}\left[(\partial_{\theta}g_{t\phi})+
\frac{2J}{M}\left(1 - \frac{\tilde{Q}}{2B_{0}J}\right)(\partial_{\theta}g_{tt})\right]
(d\theta \wedge dt) \\
  &+& \frac{1}{2}B_{0}\left[(\partial_{r}g_{\phi \phi})+
\frac{2J}{M}\left(1 - \frac{\tilde{Q}}{2B_{0}J}\right)(\partial_{r}g_{\phi t})\right]
(dr \wedge d\phi) \nonumber \\
  &+& \frac{1}{2}B_{0}\left[(\partial_{\theta}g_{\phi \phi})+
\frac{2J}{M}\left(1 - \frac{\tilde{Q}}{2B_{0}J}\right)(\partial_{\theta}g_{\phi t})\right]
(d\theta \wedge d\phi)], \nonumber
\ea
and here, $(\partial_{r}g_{t\phi})$, $(\partial_{r}g_{tt})$, $(\partial_{\theta}g_{t\phi})$,
$(\partial_{\theta}g_{tt})$, $(\partial_{r}g_{\phi \phi})$, and
$(\partial_{\theta}g_{\phi \phi})$ are as given earlier in the previous section.  

Obviously, in order to have some insight into the nature of this solution to Maxwell's equations,
one may wish to obtain the physical components of the electric field and the magnetic induction. 
This can only be achieved by projecting the Maxwell field tensor above onto an appropriate tetrad frame
like the ZAMO tetrad which, considering the Hartle-Thorne metric, can be well-defined as 
mentioned earlier. The ZAMO tetrad is well-known and is widely employed in 
various anayses in the literature. ZAMO is a fiducial observer following a timelike geodesic 
orthogonal to spacelike hypersurfaces. This implies that its 4-velocity is just the timelike
ZAMO tetrad $u^{\mu} = e^{\mu}_{0}$ given earlier in Eq. (8). Thus, if the Maxwell field 
tensor's components are projected onto
the ZAMO tetrad frame, the physical components of the electric and the magnetic fields can be read off
as $E_{i}=F_{i0}=F_{\mu\nu}(e^{\mu}_{i}e^{\nu}_{0})$ and $B_{i}=\epsilon_{ijk}F^{jk}/2=
\epsilon_{ijk}F^{\mu\nu}(e^{\mu}_{j}e^{\nu}_{k})/2$, respectively. Namely,   
from the dual to the ZAMO tetrad, $e_{A}=(e_{0}=e_{(t)}, e_{1}=e_{(r)}, 
e_{2}=e_{(\theta)}, e_{3}=e_{(\phi)})$, given earlier in Eq. (8), and
\ba
F_{10} &=& F_{\mu\nu} e^{\mu}_{1}e^{\nu}_{0} \nonumber \\
      &=& \left(\frac{\Delta}{S}\right)^{1/2}\left[(\Delta R)^{-1/2}\frac{1}{2}B_{0}
\left\{(\partial_{r}g_{t\phi})+\frac{2J}{M}\left(1 - \frac{\tilde{Q}}{2B_{0}J}\right)
(\partial_{r}g_{tt})\right\} \right.  \nonumber \\
&&\left. +(\Delta R)^{-1/2}\frac{2J}{r^3}\frac{1}{2}B_{0}
\left\{(\partial_{r}g_{\phi \phi})+\frac{2J}{M}\left(1 - \frac{\tilde{Q}}{2B_{0}J}\right)
(\partial_{r}g_{\phi t})\right\}\right], \nonumber \\ 
F_{20} &=& F_{\mu\nu} e^{\mu}_{2}e^{\nu}_{0} \nonumber \\
      &=&  \frac{1}{rA^{1/2}}\left[(\Delta R)^{-1/2}\frac{1}{2}B_{0}
\left\{(\partial_{\theta}g_{t\phi})+\frac{2J}{M}\left(1 - \frac{\tilde{Q}}{2B_{0}J}\right)
(\partial_{\theta}g_{tt})\right\} \right.  \nonumber \\
&&\left. +(\Delta R)^{-1/2}\frac{2J}{r^3}\frac{1}{2}B_{0}
\left\{(\partial_{\theta}g_{\phi \phi})+\frac{2J}{M}\left(1 - \frac{\tilde{Q}}{2B_{0}J}\right)
(\partial_{\theta}g_{\phi t})\right\}\right], \nonumber \\
F_{30} &=& F_{\mu\nu} e^{\mu}_{3}e^{\nu}_{0} = 0, \\
F_{12} &=& F_{\mu\nu} e^{\mu}_{1}e^{\nu}_{2} = 0, \nonumber \\
F_{13} &=& F_{\mu\nu} e^{\mu}_{1}e^{\nu}_{3} \nonumber \\
      &=&  \left(\frac{\Delta}{S}\right)^{1/2} \frac{1}{rA^{1/2}\sin \theta}
\frac{1}{2}B_{0}\left\{(\partial_{r}g_{\phi \phi})+
\frac{2J}{M}\left(1 - \frac{\tilde{Q}}{2B_{0}J}\right)(\partial_{r}g_{\phi t})\right\}, \nonumber \\
F_{23} &=& F_{\mu\nu} e^{\mu}_{2}e^{\nu}_{3} \nonumber \\
      &=& \frac{1}{rA^{1/2}}\frac{1}{rA^{1/2}\sin \theta}
\frac{1}{2}B_{0}\left\{(\partial_{\theta}g_{\phi \phi})+
\frac{2J}{M}\left(1 - \frac{\tilde{Q}}{2B_{0}J}\right)(\partial_{\theta}g_{\phi t})\right\}
            \nonumber                          
\ea

\begin{center}
{\rm\bf V. MAGNETIC FLUX THROUGH A NEUTRON STAR}
\end{center}

With an asymptotically uniform stationary axisymmetric magnetic field aligned
with the spin axis of a ``slightly charged'' slowly-rotating neutron star, as given above, we now would like
to compute the flux of the magnetic field across one half of the surface of the neutron star, 
which is assumed to be of exact spherical geometry with radius $r_{0}$.
The physical motivation behind this study is to have some insight into the question
of how much of the magnetic flux can actually penetrate the surface of a rotating neutron star -
at least in idealized situations. A study of this sort might seem irrelevant since for a
magnetized rotating compact object, all the flux would pass through its surface. Particularly
for pulsars (which are believed to be rotating neutron stars), carrying intrinsic magnetic dipole moments, all the
dipole magnetic field lines would be anchored on their surfaces. This naive picture, however, 
is based on electrodynamics in a {\it flat} spacetime and completely
ignores the effects of highly {\it self-gravitating} rotating neutron stars. As we demonstrated earlier,
in a region close to the surface of a rotating neutron star, Wald's solution exhibits
possible higher-multipole fields, as well as a dipole field. Moreover, there is an interesting
lesson we learned in the case of a Kerr black hole immersed in an asymptotically uniform magnetic
field, and that is the observation that if the hole is maximally-rotating, the magnetic fields
on the horizon get entirely expelled \cite{klk}. Thus, one might wonder if the same can actually 
happen in the case of rotating neutron stars. Indeed, these are the issues we would like to address 
in this section in a careful manner. 

We now begin by considering two vectors lying on the surface of
the neutron star. These vectors are given by
\be
dx_{1}^{\alpha}=(0, \; 0, \; d\theta, \; 0), \;\;\; dx_{2}^{\alpha}=(0, \; 0, \; 0, \; d\phi).
\ee
Then, in terms of a 2nd-rank tensor constructed from these two vectors \cite{hongsu},
\be
d\sigma^{\alpha\beta}=\frac{1}{2}(dx_{1}^{\alpha}dx_{2}^{\beta}-dx_{1}^{\beta}dx_{2}^{\alpha}),
\ee
one now can define the invariant surface element of the neutron star as
\ba
ds &=& (2d\sigma_{\alpha\beta}\sigma^{\alpha\beta})^{1/2}\mid_{r_{0}} \\
   &=& (g_{\theta\theta}g_{\phi\phi})^{1/2}\mid_{r_{0}} \, d\theta d\phi. \nonumber 
\ea
Next, since the tensor $d\sigma^{\alpha\beta}$ is associated with the invariant surface
element of any (not necessarily closed) 2-surface, the flux of the electric field and
the magnetic field across any 2-surfaces can be given, respectively, by
\ba
\Phi_{E} &=& \int \tilde{F}_{\alpha\beta} d\sigma^{\alpha\beta} = \int \frac{1}{2}
  \epsilon_{\alpha\beta}^{\;\;\;\;\gamma\delta}F_{\gamma\delta}d\sigma^{\alpha\beta},
  \nonumber \\
\Phi_{B} &=& \int F_{\alpha\beta} d\sigma^{\alpha\beta}.  
\ea
In particular, the flux of the magnetic field across one half of the surface of the neutron star is
\be
\Phi_{B} = \int_{r=r_{+}} F_{\alpha\beta} d\sigma^{\alpha\beta}
 = \int^{2\pi}_{0}d\phi \int^{\pi/2}_{0} d\theta F_{\theta\phi} \mid_{r_{0}}.
\ee
Then, using $F_{\theta \phi}$ evaluated on the surface of the neutron star
\small
\ba
 F_{\theta\phi} \mid_{r_{0}} &=& \frac{1}{2}B_{0}\left[(\partial_{\theta}g_{\phi \phi})+
\frac{2J}{M}\left(1 - \frac{\tilde{Q}}{2B_{0}J}\right)(\partial_{\theta}g_{\phi t})\right]  \\
&=& \frac{1}{2}B_{0}\left\{2r^2_{0} - \frac{2J}{M}\left(1-\frac{\tilde{Q}}{2B_{0}J}\right)\frac{4J}{r_{0}}\right\}
\left[1 + \frac{J^2}{Mr^3_{0}}\left(1+\frac{2M}{r_{0}}\right) \right. \nonumber \\
&&\left. - \frac{5}{8} \frac{Q-J^2/M}{M^3}\left\{
\frac{2M}{[r^2_{0} (1-2M/r_{0})]^{1/2}}Q^{1}_{2}(\frac{r_{0}}{M}-1) - Q^{2}_{2}(\frac{r_{0}}{M}-1)\right\}\right]\sin \theta \cos \theta \nonumber \\
&+& \frac{3}{2}B_{0}\left\{2r^2_{0} - \frac{2J}{M}\left(1-\frac{\tilde{Q}}{2B_{0}J}\right)\frac{4J}{r_{0}}\right\}
\left[-\frac{J^2}{Mr^3_{0}}\left(1+\frac{2M}{r_{0}}\right) \right. \nonumber \\
&&\left. + \frac{5}{8} \frac{Q-J^2/M}{M^3}\left\{
\frac{2M}{[r^2_{0} (1-2M/r_{0})]^{1/2}}Q^{1}_{2}(\frac{r_{0}}{M}-1) - Q^{2}_{2}(\frac{r_{0}}{M}-1)\right\}\right](\sin \theta \cos^3 \theta - \sin^3 \theta \cos \theta ), \nonumber 
\ea
\normalsize
and the results of integration, $\int^{\pi/2}_{0}d\theta \sin \theta \cos \theta = 1/2$ and
$\int^{\pi/2}_{0}d\theta (\sin \theta \cos^3 \theta - \sin^3 \theta \cos \theta) = 0$,
we finally arrive at
\ba
&&\Phi_{B} = \pi B_{0}\left\{r^2_{0} - \frac{4J^2}{Mr_{0}}\left(1-\frac{\tilde{Q}}{2B_{0}J}\right)\right\}\times  \\
&&\left[1 + \frac{J^2}{Mr^3_{0}}\left(1+\frac{2M}{r_{0}}\right) 
- \frac{5}{8} \frac{Q-J^2/M}{M^3}\left\{
\frac{2M}{[r^2_{0} (1-2M/r_{0})]^{1/2}}Q^{1}_{2}(\frac{r_{0}}{M}-1) - Q^{2}_{2}(\frac{r_{0}}{M}-1)\right\}\right] .
\nonumber
\ea
This is the magnetic flux through one half of the surface of a slowly-rotating neutron star with
accretion electric charge $\tilde{Q}$. Some discussions concerning the nature of this magnetic flux
are now in order :  \\

(i) In the absence of accreted charge, i.e., $\tilde{Q}=0$, the magnetic flux above appears
to drop to zero (i.e., the magnetic field gets entirely expelled) when the neutron star is
maximally-rotating, which, based on the expression for the magnetic flux across the surface of
the neutron star given in Eq. (39), amounts to $J^2_{max} = Mr^3_{0}/4$. This curious feature, indeed,
has a counterpart in the Kerr black hole case \cite{klk}, and that counterpart has been known for some time.
For the present case of a neutron star, however, we are not fully qualified to make any definite
statement on this point because the Hartle-Thorne metric employed represents the region surrounding a
{\it slowly-rotating} neutron star. Thus, the expression for the magnetic flux given above may be
invalid for the case of rapidly-rotating neutron star, but that is beyond the scope of the
present work as the associated metric is not available for now. \\

(ii) For a nonvanishing accreted charge having a value in the range $0<\tilde{Q} <2B_{0}J$,
it is interesting to note that
\begin{eqnarray}
\Phi_{B}(\tilde{Q}\neq 0) > \Phi_{B}(\tilde{Q}= 0).
\end{eqnarray}
Actually, this is one of the points of central importance we would
like to make in the present work. The physical interpretation of this characteristic
can be briefly stated as follows : Once the neutron star accretes charge, toroidal currents occur
as a result of rotation of the star. These toroidal currents, in turn, generate new magnetic  
fields which would be additive to the existing ones, increasing the total magnetic flux through
the surface of the rotating neutron star. As in Ref. 9,
essentially the same phenomenon takes place in the case of a slightly charged Kerr black hole.
There, the interpretation was even clearer ; when the spin of the hole and the asymtotically
uniform magnetic field are parallel (antiparallel), the hole selectively accretes
positive (negative) charge according to the injection energy argument proposed by Carter, and 
the charge, in turn, generates magnetic fields which are clearly additive to the existing ones. 
Also we expect that essentially the same will be true for the present rotating neutron star case
although we could not determine the sign of the accreted charge in association with the relative
direction between the spin of the neutron star and the direction of the magnetic field due to
the highly complex structure of Hartle-Thorne metric, as we mentioned earlier. \\

(iii) For the Wald charge $\tilde{Q}=2B_{0}J$, the magnetic flux across the surface of the
rotating neutron star appears to take on a rather special value :
\ba
&&\Phi_{B} = B_{0}\pi r^2_{0}\times  \\
&&\left[1 + \frac{J^2}{Mr^3_{0}}\left(1+\frac{2M}{r_{0}}\right) 
- \frac{5}{8} \frac{Q-J^2/M}{M^3}\left\{
\frac{2M}{[r^2_{0} (1-2M/r_{0})]^{1/2}}Q^{1}_{2}(\frac{r_{0}}{M}-1) - Q^{2}_{2}(\frac{r_{0}}{M}-1)\right\}\right] .
\nonumber
\ea
As mentioned in the introduction, it may be interesting to compare this analysis of the nature of
the magnetic flux through a rotating neutron star with that of the Kerr hole studied in Ref. 9.
That is, for all three cases (i), (ii), and (iii), the results discussed are generally {\it consistent} with those
in the Kerr black hole case although the quantitative details are rather different between the two cases,
the neutron star and the black hole, as the spacetime metrics representing the two are distinct.

\begin{center}
{\rm\bf VI. CONCLUDING REMARKS}
\end{center}
  
In the present work, based on the solution-generating method given by Wald, the magnetic fields around
both uncharged and (slightly) charged neutron stars have been obtained. Particularly, for the
charged neutron star, it has been demonstrated, following again the argument by Wald, that 
the neutron star will gradually accrete charge until it reaches an equilibrium value 
$\tilde{Q}=2B_{0}J$. In association with the ``magnetic braking'' model of Goldreich and Julian \cite{gj}
for the spin-down of magnetized rotating neutron stars, the magnetic flux through one half of the
surface of the rotating neutron star was computed, as well. For a nonvanishing accretion 
charge having a value in the range $0 < \tilde{Q} \leq 2B_{0}J$, the total magnetic flux through the 
neutron star has been shown to be greater than that without the accreted charge.
Indeed, the physical interpretation of this characteristic
can be briefly stated as follows : When the spin of the neutron star and the direction of the asymtotically
uniform magnetic field are parallel (antiparallel), the neutron star will selectively accrete
positive (negative) charges following the injection energy argument proposed by Carter, and 
those charges, in turn, generate magnetic fields that add to the existing ones.   

Although our discussion thus far is equally relevant to all species of slowly-rotating relativistic stars,
indeed, we particularly had rotating neutron stars in mind. As such, it would be
natural to attempt to make contact with the case of real pulsars as they are believed to 
be magnetized rotating neutron stars. Thus, below we consider how many of the theoretical results obtained in the
present work can be carried over to a realistic, general relativistic description of the pulsar's
magnetosphere. 

Radio pulsars are perhaps the oldest-known and the least energetic among all of the pulsar categories.
The thoretical study of these radio pulsars can be traced back to the 1969 work of Goldreich and 
Julian \cite{gj}. In their pioneering work, Goldreich and Julian argued that pulsars, which are
thought to be the rotating magnetized neutron stars, must have a {\it magnetosphere} with a
charge-separated plasma. They then demonstrated that an electric force, which is much
stronger than the gravitational force, will be set up along the magnetic field and as a result, 
the surface charge layer cannot be in dynamical equilibrium. Then, there appear steady current flows 
along the magnetic field lines, which are taken to be uniformly rotating since they are firmly rooted 
in the crystalline crust of the pulsar surface. Although it was rather implicit in their original work,
this model for the pulsar electrodynamics in terms of the force-free magnetosphere does, indeed,
suggest that the luminosity of radio pulsars is due to the loss of their rotational energy (namely, 
the spin-down) and its subsequent conversion into charged particle emission, which eventually 
generates radiation in the far zone.

Now, it is our hope to relate the results obtained in the present work to the physics of real pulsars ;
thus, we begin with an inherent drawback of our treatment of the issues presented in this work.
In the present work, we have mainly considered the case with a symmetric geometry in which the 
stationary axisymmetric magnetic field is precisely aligned with the neutron star's axis of rotation.
Although this simplification/idealization was inevitable in order to make the system mathematically
tractable, the magnetized rotating neutron star will fail to describe pulsars
in the real world as the pulsars involve, by definition, oblique spin-magnetic field configurations.
(In the case involving a rotating black hole, however, such an oblique configuration cannot be
maintained because any possible initial misalignment between the spin and the direction of the
magnetic field will be removed in an efficient manner either by a viscous torque when the accretion
disc is involved \cite{bp} or by a magnetic alignment torque \cite{align}.)
Indeed, almost all of the theoretical studies of pulsar electrodynamics assume this idealized
symmetric geometry for concrete, analytical calculations, and concerning this limitation,
our work makes no progress. Thus we are still far away from a satisfactory 
theoretical description of the real physics that takes place in the pulsar's magnetosphere.
Nevertheless, there are quite a few features represented by the theoretical results in the present work
that could be relevant to the real pulsar situation. To make this point indeed look convincing, 
we first comment on our motivation for choosing the Wald algorithm to
obtain the solutions of Maxwell's equations in the background spacetime of neutron stars.
Once again, the philosophy employed in the present work is to take the electromagnetic fields 
surrounding a rotating neutron star as a theoretical model to represent the pulsar magnetosphere
(except possibly for the plasma content). Therefore, as a {\it theoretical} model, our approach
is, indeed, not without some inherent limitations. We here remind the reader that earlier in the 
introduction, we mentioned that the results presented here for the case of a rotating
neutron star can be compared in parallel with those that have been presented in our earlier 
work \cite{hongsu} for the case of a rotating Kerr black hole because we employed in this work 
the same algorithm of Wald to determine the structure of (electro)magnetic field around the 
rotating neutron star. Now, considering this point, notice that in Wald's construction of a
stationary axisymmetric solutions of Maxwell's equations in a Kerr black hole spacetime or
in a Hartle-Thorne spacetime for a rotating neutron star, we assumed on technical grounds
from the outset that the Kerr black hole or the rotating neutron star was placed in an originally
uniform magnetic field. Indeed, this setup could be relevant for objects like pulsars which
produce very strong external dipole magnetic fields. After all, this asymptotically uniform
magnetic field configuration is, at least, {\it not} inconsistent with the open dipole magnetic
field lines in the far zone in the ``aligned rotator'' model of Goldreich and Julian \cite{gj}.
Then, one looks for a surrounding (electro)magnetic field, which gets distorted by the presence
of a Kerr black hole or a rotating neutron star. In a rigorous sense, therefore, this boundary 
condition in Wald's construction may not seem very relevant for real pulsar situations in which 
pulsars are believed to carry dipole moments and, hence, produce dipole magnetic fields.
In this regard, our approach based on Wald's construction method can be viewed as one seeking
systematic gravitational corrections to an intrinsic dipole magnetic field. Indeed, as we
stated earlier, Wald's solution has no monopole moment. In the region close to a neutron
star, however, it may well have a non-trivial {\it dipole} and possibly higher mutipole fields.
Also, we believe that this behavior of Wald's solution is a new contribution to the existing
information on the structure of a pulsar magnetosphere given in the literature. 

Lastly, one might be worried about the validity of the Hartle-Thorne metric for the region 
surrounding the slowly-rotating neutron stars employed in this work to describe the magnetosphere 
of pulsars, which seem to be rapidly-rotating and to have typically millisecond pulsation periods.
Thus, in the following, we shall defend this point in a careful manner.
Here, ``slowly-rotating'' means that the neutron star rotates relatively slowly compared to
an equal-mass Kerr black hole which can rotate arbitrarily rapidly up to its maximal rotation of
$J = M^2$. Thus, this does not necessarily mean that the Hartle-Thorne metric for slowly-rotating
neutron stars cannot properly describe millisecond pulsars. To see this, note that according
to the Hartle-Thorne metric, the angular speed of a rotating neutron star is given by the 
Lense-Thirring precession angular velocity in Eq. (9) at the surface of the neutron star, which,
restoring the fundamental constants to get back to Gaussian units, is 
\ba
\omega = \frac{2J}{r^3_{0}}\left(\frac{G}{c^2}\right), ~~~{\rm with}
~~J = \tilde{a}M^2\left(\frac{G}{c}\right) ~~(0<\tilde{a} <1).
\ea
As we mentioned earlier, one of the obvious differences between the black hole case and the neutron star
case is the fact that the black hole is characterized by its event horizon while the neutron
star has a hard surface. As such, in terms of the spacetime metric generated by each of them, just
as the Lense-Thirring precession angular velocity (due to frame-dragging) at the horizon represents
the black hole's angular velocity, the Lense-Thirring precession angular velocity at the location
of the neutron star's surface should give the angular velocity of the rotating neutron star. 

Thus, the Hartle-Thorne metric gives the angular speed of a rotating neutron star having the 
parameters of a typical radio pulsar, $M\sim 2\times 10^{33}(g)$ and $r_{0}\sim 10^6 (cm)$, as
$\omega = 2\tilde{a}(M^2/r^3_{0})(G/c)(G/c^2) \sim 10^3 (1/sec)$ which, in turn, yields a
rotation period of $\tau = 2\pi/\omega \sim 10^{-2} (sec)$. Here we used
$(G/c^2) = 0.7425\times 10^{-28} (cm/g)$ and $(G/c) = 2.226\times 10^{-18} (cm^2/g\cdot sec)$.
Indeed, this is impressively
comparable to the observed pulsation periods of radio pulsars, $\tau \sim 10^{-3} - 1 (sec)$, as
we discussed earlier. As a result, we expect the Hartle-Thorne metric to be well-qualified to
describe the geometries of millisecond pulsars. 

To summarize, some aspects of the results of the theoretical study of a rotating neutron star's
electrodynamics carried out in this work seem irrelevant while some appear to be quite relevant 
to real pulsar situations. Namely, although the predictions of the present theoretical study
of a rotating neutron star's electrodynamics are not fully satisfying, it is, nevertheless, our
hope that at least here we have taken one step closer to a systematic general relativistic
study of the electrodynamics in a region close to a rotating neutron star in association
with their pulsar interpretation.

\vspace*{1cm}

\begin{center}
{\rm\bf Acknowledgements}
\end{center}

H. Kim would like to thank Prof. L. Rezzolla for relevant comments on the electrodynamics of
neutron stars via a private communication.
H. M. Lee was supported by a Korean Research Foundation grant (No. D00268) in 2001.

\end{document}